# A Fuzzy Differential Evolution Algorithm for Job Scheduling on Computational Grids

Ch.Srinivasa Rao[1], Dr.B.Raveendra Babu[2]

[1]*Assoc.Professor, Dept.of Computer Applications, RVR & JC College of Engineering*
*Guntur, INDIA*

[2]*Professor, VNR Vignana Jyothi Institute of Engineering & Technology,*
*Hyderabad, INDIA*

*Abstract*- Grid computing is the recently growing area of computing that share data, storage, computing across geographically dispersed area. This paper proposes a novel fuzzy approach using Differential Evolution (DE) for scheduling jobs oncomputational grids. The fuzzy based DE generatesan optimal plan to complete the jobs within a minimum period of time. We evaluate the performance of the proposed fuzzy based DE algorithm with GeneticAlgorithm (GA), Simulated Annealing (SA), Differential Evolution and fuzzy PSO. Experimental results have shown that the new algorithm produces more optimal solutions for the job scheduling problems compared to other algorithms.

## I. INTRODUCTION

Grid computing is the recently growing area of computing that share data, storage, computing across geographically dispersed area [1]. Grid resources are shared by submitting computing tasks to grid system. The resources of computational grid are dynamic and belongtodifferent administrative domains. The participation of resources can be active or inactive within the grid.It is impossible to assign jobs manually to computing resources in grids. Thus the gridjob scheduling is became one of the challenging issues in grid computing. Grid scheduling system selects theresources and allocates the user submitted jobs to appropriate resources in such a way that the user andapplication requirements are met.Grid computing provides competitive solutions to complex tasks in shorter time and utilizes the hardware efficiently. The efficient job scheduling methods are required to achieve high performance in grid computing environment. Job scheduling is the essential step in grid computing where the jobs are scheduled to different machines. There exits different job scheduling methodologies using centralized scheduling, distributive scheduling, hierarchical scheduling, agent based scheduling, dynamic job scheduling etc[2]. However, it is abig challenge to develop the best scheduling algorithm. The problem isbecomemuch more complex today with added dynamic nature, heterogeneity of jobs and resources of Grid systems. Therefore the use of meta- heuristic optimization approaches became vital for job scheduling problems.

The difficulty in optimization of engineering problems have initiates the researchers to find various optimization algorithms. As a result several heuristic algorithms are developed for optimization of parameters. Among these one important group is evolutionary algorithms (EA). Some of the Evolutionary Algorithms are Genetic Algorithms (GA), Evolution Strategy (ES), Evolution Programming (EP) and Differential Evolution (DE). The most commonly used evolutionary optimization technique is the Genetic Algorithm (GA). Though, the GA provides a near optimal solution for complex problems, is require number of control parameters in advance such as crossover rate and mutation rate, which affect the effectiveness of the solution. Determining the optimum values for these controlling parameters is very difficult in practical. Differential Evolution (DE) is one of the most powerful stochastic real-parameter optimization algorithms in current use [3] [4] [5] [6] [7]. DE follows similar computational steps as in a standard evolutionary algorithm. DE uses a weighted difference of the solution vectors to explore the objective function in population. Compared to other Evolutionary Algorithms DE is very simple to code [8]. The recent studies on DE have shown that DE provides a better performance in terms of accuracy, robustness and convergence speed with its simplicity [9]. The number of control parameters in DE is very few compared to other algorithms. DE is became a successful technique for many applications [10] [11]. The DE algorithm is also extended as a competitive solution for various multi objective problems [12][13].

Krauter et al. provided a useful survey on grid resource management systems, in which most of the grid schedulers such as AppLes, Condor, Globus, Legion, Netsolve, Ninf and Nimrod use simple batch scheduling heuristics [14]. Jarvis et al. proposed the scheduling algorithm using metaheuristics and compared FCFS with genetic algorithm to minimize the makespan and it was found that metaheuristics generate good quality schedules than batch scheduling heuristics [15]. Braun et al. studied the comparison of the performance of batch queuing heuristics, tabu search, genetic algorithm and simulated annealing to minimize the makespan [16]. The results revealed that genetic algorithm achieved the best results compared to batch queuing heuristics. Hongbo Liu et al. proposed a fuzzy particle swarm optimization (PSO) algorithm for scheduling jobs on computational grid with the minimization of makespan as the main criterion [17]. They empirically showed that their method outperforms the genetic algorithm and simulated annealing approach. The results revealed that the PSO algorithm has an advantage of high speed of convergence and the ability to obtain faster and feasible schedules. Recently Srinivasa Rao and Raveendra Babu developed a DE based solution for job scheduling algorithms[18]. As Fuzzy provides more prominent solutions compared to hard approaches, we have developed a new fuzzy based DE for the job scheduling problems. This paper proposes the fuzzy based DE and evaluates the performance of the proposed with four different data sets varying size and capacity. The experimental results showed the improved performance of the proposed algorithm.

## II. SCHEDULING PROBLEM

Scheduling is the process of mapping the jobs to specific time intervals of the grid resources. The grid job scheduling problem consists of scheduling m jobs with given processing time on n resources. Let Jj be the independent user jobs, j = {1, 2, 3…m}. Let Ribe the heterogeneous resources, i = {1, 2, 3…n}. The speed of each resource is expressed in number of cycles per unit time (CPUT). The length of each job is expressed in number of cycles. The information related to job length and speed of the resource is assumed to be known, based on user supplied information, experimental data and application profiling or other techniques [19].

The objective of the proposed job scheduling algorithm is to minimize the makespan. Makespan is a measure of the throughput of the heterogeneous computing system. Let $C_{i,j}$ ($i \in \{1,2,...n\}$, $j \in \{1,2,...m\}$) be the completion time that the resource Ri finishes the job Jj, $\sum C_i$ represents the time that the resource Ri finishes all the jobs scheduled for itself. Makespan is defined as $C_{max} = \max \{\sum C_i\}$ [20].

## III. DIFFERENTIAL EVOLUTION PROCEDURE

### A. Initialization

Creation of a population of individuals. The $i^{th}$ individual vector (chromosome) of the population at current generation $t$ with $d$ dimensions is as follows,

$$Z_i(t) = [Z_{i,1}(t), Z_{i,2}(t), \cdots, Z_{i,d}(t)] \quad (2)$$

### B. Mutation

For each individual vector $Z_k(t)$ that belongs to the current population, a new individual, called the mutant individual is derived through the combination of randomly selected and pre-specified individuals.

$$U_{k,n}(t) = Z_{m,n}(t) + F * (Z_{i,n}(t) - Z_{j,n}(t)) \quad (3)$$

the indices m, n, i, j are uniformly random integers mutually different and distinct from the current index k, and F > 0 is a real positive parameter, called mutation or scaling factor (usually∈[0, 1]).

### C. Recombination (Crossover)

DE has two crossover schemes: the exponential and the binomial or uniform crossover. We have used the binomial crossover in this paper. The binomial or uniform crossover is performed on each component $n$ (n= 1, 2, . . . , d) of the mutant individual $U_{k,n}(t)$. For each component a random number $r$ in the interval [0, 1] is drawn and compared with the crossover rate or recombination factor (another DE control parameter), CR ∈ [0, 1]. If r <=CR, then the nth component of the mutant individual $U_{k,n}(t)$ will be selected, Otherwise, the $n^{th}$ component of the target vector $Z_{k,n}(t)$ becomes the $n^{th}$ component of the trial vector.

$$U_{k,n}(t+1) = \begin{cases} U_{k,n}(t), & \text{if } rand_n(0,1) < CR \\ Z_{k,n}(t), & \text{otherwise} \end{cases} \quad (4)$$

### D. Selection

Choice of the best individuals for the next cycle. If the new offspring yields a better value of the objective function, it replaces its parent in the next generation; otherwise, the parent is retained in the population, i.e.,

$$Z_k(t+1) = U_k(t+1), \quad \text{if } f(U_k(t+1)) > f(Z_k(t))$$
$$Z_k(t), \quad \text{if } f(U_k(t+1)) < f(Z_k(t)) \quad (5)$$

Where f(·) is the objective function to be minimized. In this paper we have used makespan as the objective function.

## IV. FUZZY DE FOR JOB SCHEDULING

In this section, we propose a fuzzy based DE to solve the job scheduling problem on computational grids. The solution or each chromosome is a matrix that represents allocation of jobs to resources. Assume that the resources are R={ $R_1, R_2,…,R_m$} and Jobs to allocate are J={$J_1,J_2,…,J_n$}, then the fuzzy scheduling relation is as follows:

$$\text{Membership matrix (F)} = \begin{pmatrix} F_{11} & F_{12} & \cdots & F_{1n} \\ F_{21} & F_{22} & \cdots & F_{2n} \\ \vdots & \vdots & \cdots & \vdots \\ F_{m1} & F_{m2} & \cdots & F_{mn} \end{pmatrix}$$

Where $F_{ij}$ represents the degree of membership of the $i^{th}$ Resource to the $j^{th}$ Job. The fuzzy relation F between R and J has the following meaning: for each element in the matrix F, the element.

$$F_{ij} = \mu_R(R_i, J_j), \quad i \epsilon \{1,2,…,m\}, \quad j \epsilon \{1,2,…,n\}. \quad (6)$$

$\mu_R$ is the membership function, the value of $s_{ij}$ means the degree of membership that the grid node $G_j$ would process the job $J_i$ in the feasible schedule solution. In the grid job scheduling problem, the elements of the solution must satisfy the following conditions:

$$F_{ij} \in [0,1], i \in \{1,2,…,m\}, j \in \{1,2,…,n\}. \quad (7)$$

$$\sum_{i=1}^{m} F_{ij} = 1, i \in \{1,2,\ldots,m\}, j \in \{1,2,\ldots,n\}. \quad (8)$$

The pseudo code for DE based grid job scheduling algorithm is illustrated in Algorithm 1. Table 1 depicts the explanation of abbreviated parameters used in Algorithm 1.

**Algorithm 1 Grid Job Scheduling using Fuzzy DE**

**Define** *RT, JT, ESR, JL, F, CR, NP, MaxIter, STR, ETR*
Create the initial population using Fuzzy concept of random individuals, where each individual is a fuzzy matrix.

**for** 1 to *MaxIter*
    Calculate the makespan of each individual, by determining allocation using highest membership
  **for** *i = 1 to NP*
  Select random integer $randn_i \in (0, 1, 2\ldots JT)$
  Select mutually exclusive random individuals $X_a, X_b, X_c$
  Calculate mutant vector V according to equation (2) starting from the position $randn_i$ of each individual.
  Select the random value $rand_j \in [0, 1]$
  Calculate the trail vector $U_i$ according to equation (3)
  Check the feasibility of trail vector $U_i$
  **end for**
  Calculate the makespan of trail vector set by determining allocation using highest membership
  **for** *i = 1 to NP*
   **if** makespan of $U_i$ is less than $X_i$ **then** Select $U_i$
   **else** Retain $X_i$
   **end if**
  **end for**
Record the solution with minimum makespan
**end for**

TABLE 1 PARAMETERS USED IN ALGORITHM 1

| RT | Total Resources | CR | Crossover Factor |
|---|---|---|---|
| JT | Total Jobs | NP | Population Size |
| ESR | Execution speed of Resource | MaxIter | Maximum number of Iteration |
| JL | Job length | STR | Start time of resource engaged in grid |
| F | Scaling factor | ETR | End time of resource engaged in grid |

## V. EXPERIMENTAL RESULTS AND DISCUSSIONS

Experiments are conducted using four different data sets with varying sizes. We have run hundred times the algorithm for each data set. Table2 reports the average makespan of hundred runs from various algorithms for different resource job pairs. Similarly Table3 and Table4 demonstrates the time required in seconds to converge the solution in a single run and standard deviation of the makespan in hundred iterations.

TABLE 2: PERFORMANCE COMPARISON OF THE FOUR ALGORITHMS USING THE PARAMETER MAKESPAN

| Algorithm | Resource Job Pair | | | |
|---|---|---|---|---|
| | (3,13) | (5,100) | (8,60) | (10,50) |
| GA | 47.1167 | 85.7431 | 42.9270 | 38.0428 |
| SA | 46.6000 | 90.7338 | 55.4594 | 41.7889 |
| Fuzzy PSO | 46.2667 | 84.0544 | 41.9489 | 37.6668 |
| DE | 46.0500 | 86.0138 | 43.0413 | 37.5748 |
| Fuzzy DE | 46.0166 | 85.5431 | 41.7580 | 36.05588 |

TABLE 3: PERFORMANCE COMPARISON OF THE FOUR ALGORITHMS WITH THE TIME OF COMPLETION IN SECONDS

| Algorithm | Resource Job Pair | | | |
|---|---|---|---|---|
| | (3,13) | (5,100) | (8,60) | (10,50) |
| GA | 302.9210 | 2415.9000 | 2263.0000 | 2628.1000 |
| SA | 332.5000 | 6567.8000 | 6094.9000 | 6926.4000 |
| Fuzzy PSO | 106.2030 | 1485.6000 | 1521.0000 | 1585.7000 |
| DE | 81.5203 | 435.8865 | 337.7940 | 346.3016 |
| Fuzzy DE | 114.8011 | 424.4141 | 464.7304 | 365.2094 |

TABLE 3: PERFORMANCE COMPARISON OF THE FOUR ALGORITHMS WITH THE STANDARD DEVIATION IN 100 RUNS

| Algorithm | Resource Job Pair | | | |
|---|---|---|---|---|
| | (3,13) | (5,100) | (8,60) | (10,50) |
| GA | 0.7700 | 0.6217 | 0.4150 | 0.6613 |
| SA | 0.4856 | 6.3833 | 2.0605 | 8.0773 |
| Fuzzy PSO | 0.2854 | 0.5030 | 0.6944 | 0.6068 |
| DE | 0.2916 | 0.3146 | 0.5274 | 0.6722 |
| Fuzzy DE | 0.096225 | 0.009181 | 0.142586 | 0.347058 |
| fuzzyGA | 0.28810 | 0.022874 | 0.031299 | 0.327824 |

The observed makespan of fittest individual in each of hundred runs is plotted in the following figures. Figure1 contains resource job pair (3,13) makespan, Figure2 reports ( 5,100), Figure3 plots (10,50) and Figure4 displays (8,60) makespan in various runs.

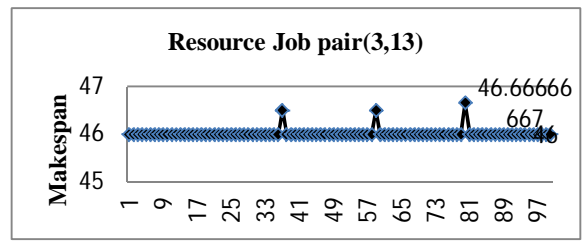

Fig.1 Makespan of (3,13)

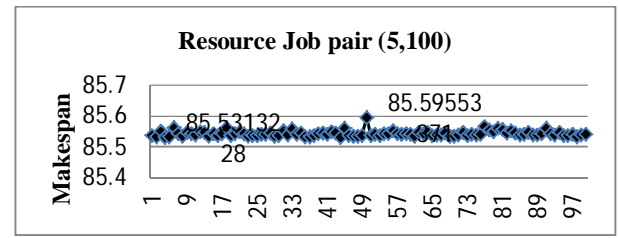

Fig.2 Makespan of (5,100)

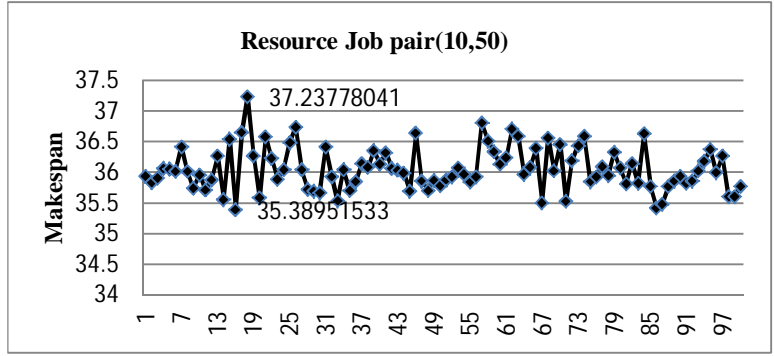

Fig.3 Makespan of (10,50)

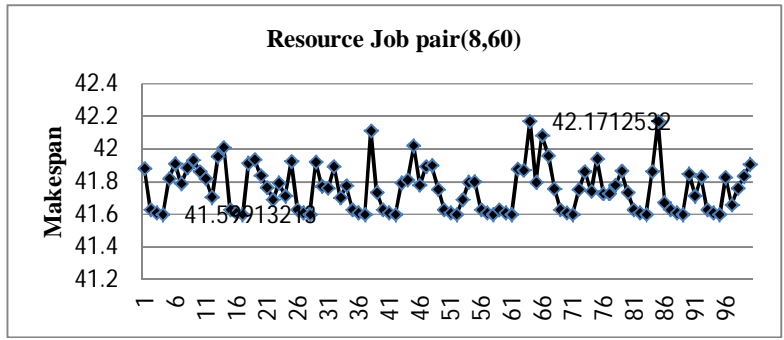

Fig.4 Makespan of (8,60)

TABLE 4: RELATIVE PERFORMANCE OBSERVED WITH VARIOUS ALGORITHMS

| Algorithm | (3,13) | (5, 100) | (8,60) | (10,50) | Average |
|---|---|---|---|---|---|
| GA | 1.10003 | 0.19994 | 1.16895 | 1.98692 | 1.11396 |
| SA | 0.58333 | 5.19064 | 13.70135 | 5.73302 | 6.302085 |
| Fuzzy PSO | 0.25003 | -1.48876 | 0.19085 | 1.61092 | 0.14076 |
| DE | 0.03333 | 0.47064 | 1.28325 | 1.51892 | 0.826535 |

Table 4 depicted that overall improvement of Fuzzy DE over GA in all cases is 1.11, over SA is 6.3, over Fuzzy PSO is 0.14 and over DE observed that 0.82. Fuzzy DE is equally performs well with Fuzzy PSO and reported in Fig.9.

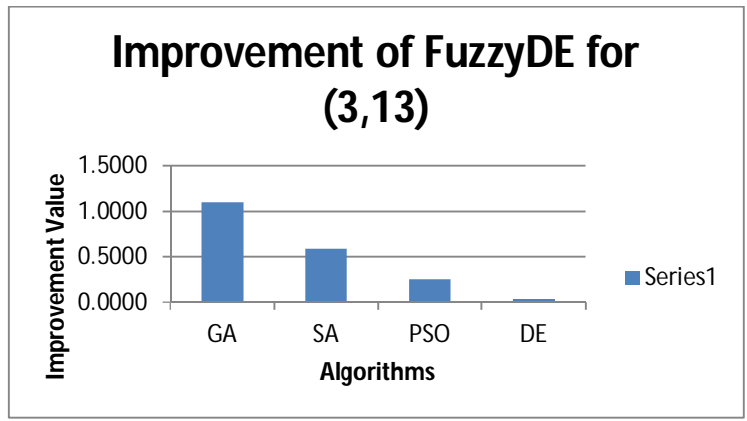

Fig.5 Improvable performance of Fuzzy DE for (3,13)

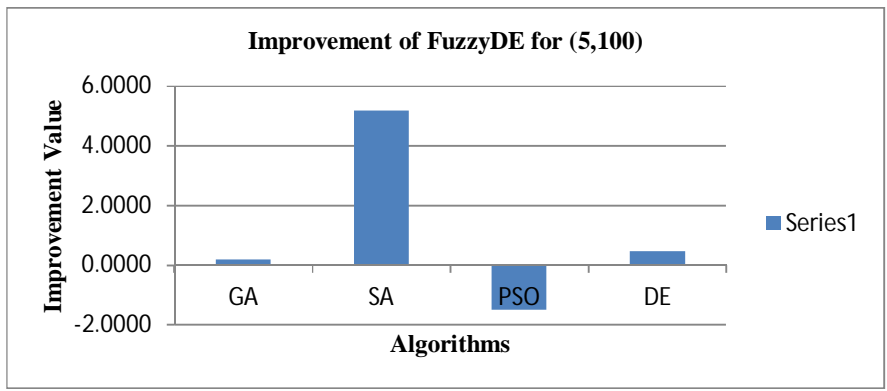

Fig.6 Improvable performance of Fuzzy DE for (5,100)

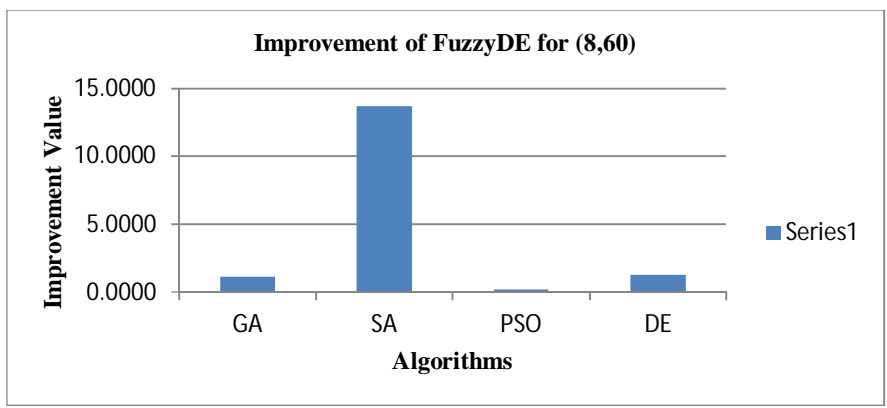

Fig.7 Improvable performance of Fuzzy DE for (8,60)

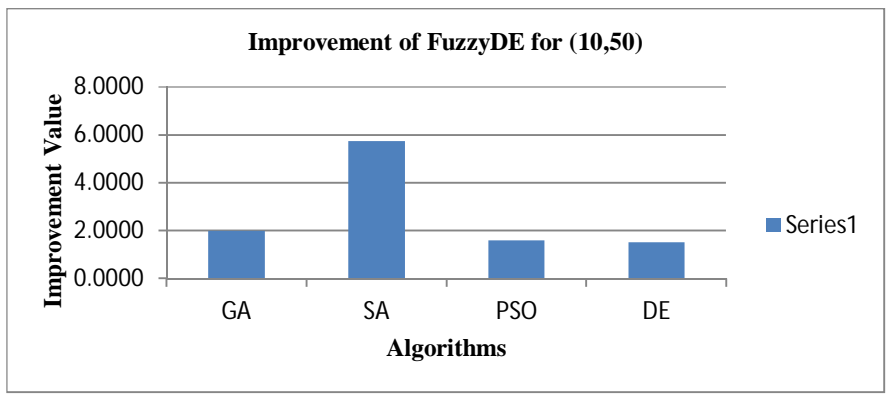

Fig.8 Improvable performance of Fuzzy DE for (10,50)

We have extended our study by reporting improved performance of Fuzzy DE towards makespan over other algorithms. From the Fig.5, GA exhibits least performance for resource job pair (3,13). Fig.6 demonstrates that SA has shown highest makespan whereas Fuzzy PSO is better than Fuzzy DE by 1.48876. The improvable performance of Fuzzy DE is more in the case of resource job pair (8,60) and it is equally performs well with Fuzzy PSO. From the Fig.8, it is observed that GA, Fuzzy PSO and DE are equally performed well and approximately Fuzzy DE improvement is 1.5.

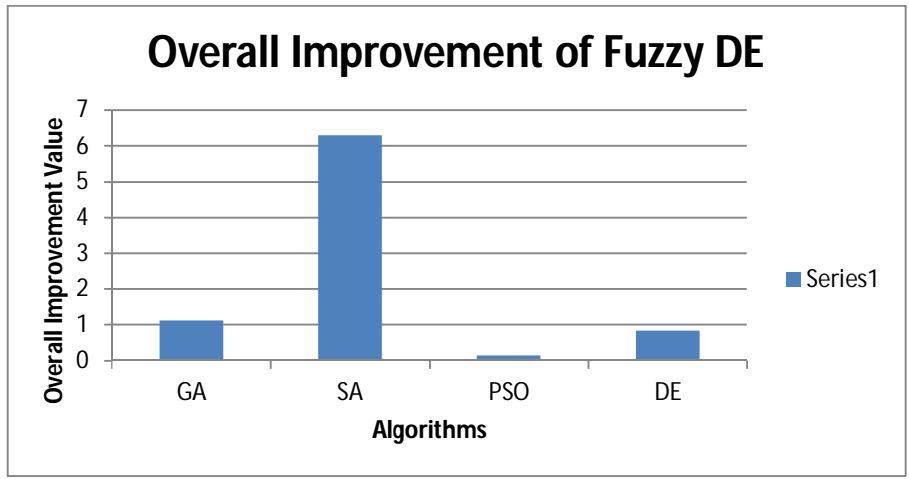

Fig.9 Relative performance of Fuzzy DE

## VI. CONCLUSIONS

The proposed Fuzzy DE has been developed incorporating fuzzy logic in Differential Evolution algorithm. The performance of Fuzzy DE is studied using various data sets and compared with various other evolutionary algorithms. The experimental results have shown that Fuzzy DE reported optimal solution in each case towards makspan. From the observation, Fuzzy DE is equally good with Fuzzy PSO developing a new algorithm which provides more optimal solutions in future endeavour. In our future study, we will consider the problems including the processing time.

ACKNOWLEDGMENT

We would like to thank, Professor K.Karteeka Pavan for supporting the work proposed in this paper.